# Do journals flipping to Gold Open Access show an OA Citation or Publication Advantage?

Nuria Bautista-Puig[1], Carmen Lopez-Illescas [2], Felix de Moya-Anegon[3], Vicente Guerrero-Bote[4], Henk F. Moed*[5]

[1]nbautist@bib.uc3m.es.
University Carlos III of Madrid. Research Institute for Higher Education and Science (INAECU), Getafe Madrid.

[2]*carmlopz@gmail.com*
University Complutense of Madrid. Information Science Faculty. Dept. Information and Library Science, SCImago Group, Madrid, Spain

[3]*felix.moya@scimago.es*
SCImago Group, Madrid, Spain

[4]*guerrero@unex.es*
SCImago Group, Dept. Information and Communication, University of Extremadura, Badajoz, Spain

[5]*henk.moed@uniroma1.it*
Sapienza University of Rome, Italy

* Corresponding author

## Summary

The effects of Open Access (OA) upon journal performance are investigated. The key research question holds: How does the citation impact and publication output of journals switching ("flipping") from non-OA to Gold-OA develop after their switch to Gold-OA? A review is given of the literature, with an emphasis on studies dealing with flipping journals. Two study sets with 119 and 100 flipping journals, derived from two different OA data sources (DOAJ and OAD), are compared with two control groups, one based on a standard bibliometric criterion, and a second controlling for a journal's national orientation. Comparing post-switch indicators with pre-switch ones in paired T-tests, evidence was obtained of an OA *Citation* advantage but *not* of an OA *Publication* Advantage. Shifts in the affiliation countries of publishing and citing authors are characterized in terms of countries' income class and geographical world region. Suggestions are made for qualitative follow-up studies to obtain more insight into OA- or reverse OA-flipping.

## Introduction

*Introduction to the research presented in this paper*

The starting point of the open access (OA) movement is the Budapest Open Access Initiative in 2002 (BOAI, 2002). Open access of research outputs means that such outputs are distributed online, free of cost and free of other access barriers. One can distinguish two main business models of scientific publishing, denoted concisely as a “readers pay” model and an “authors pay” model of journal publishing. The first allows access to a journal’s content on the basis of a payed subscription, while the second requires that publication costs are being paid prior to publication by the authors, or their institutions, research funders or subsidizing organizations. The terms “closed” or “pay-walled” are often used in relation to a subscription based journal publication model. Expanding the use of these terms, it can be stated that subscription-based access is closed or pay-walled towards *readers*, and open access towards publishing *authors*.

Insight into the characteristics of these two publishing models, and into their effects upon the quality of the scientific-scholarly communication system, is of the utmost importance, both for researchers, publishers, policy makers, and other interested audiences. It is therefore not surprising that these issues are among the most important research subjects in the field of information science, bibliometrics and informetrics.

The current paper examines the effects of Open Access as well. It focuses on one particular OA model, denoted as *Gold* Open Access, according to which a publisher makes *all* articles and related content in a journal available for free immediately on the journal's website. It analyses how journals that switch from a non-OA to a Gold-OA model develop after switching to Gold OA, compared to their situation before the switch. A switch from non-OA to Gold-OA is denoted by Solomon, Laakso & Björk (2016) as flipping. The approach adopted in the current study is bibliometric: it compares the citation impact of a journal after the flip with that made before the conversion to Gold-OA.

*Measuring OA: data sources*

Which data sources are available that provide information on the OA status of journals or individual articles? The Directory of Open Access Journals (DOAJ) is a website that hosts a community-curated list of open access journals. “The project defines open access journals as scientific and scholarly journals making *all* their content available for free, without delay or user-registration requirement, and meeting high quality standards, notably by exercising peer review or editorial quality control” (DOAJ, n.d.). Journals making all their content available for free are denoted as gold open access journals. For more information about the ideas behind DOAJ the reader is referred to Morrison (2008).

As a rule, DOAJ does not include hybrid open access journals. A hybrid open-access journal is a subscription-based journal in which some of the articles are open access. The open access status for an article can be acquired by paying to the publisher an article processing charge (APC), in addition to the payment of a subscription to all other articles published in the journal. Dallmeier-Tiessen et al. (2010) give more information on models and attribute of Open Access publishing.

Sotudeh & Horri (2007) presented one of the first extensive validations of data sources on OA. They studied the history of the adoption of OA by a given journal, using data from different data sources. A key piece of information was the data at which a journal switched to Gold-OA. They checked the data using the Wayback Machine tool, also known as Internet Archive, and, in case of doubt, they contacted the journal publisher or editor. They concluded that “OA directories, especially the DOAJ did not provide sufficient information in this regard” (Sotudeh and Horri, 2007, p. 1580).

In a recent study, Akbaritabar & Stahlschmidt (2019) examined the data quality and usefulness of a series of data sources on OA, including DOAJ (see www.doaj.org), Unpaywall (a database covering OA content from thousands of scientific publishers and repositories – see www.unpaywall.org); Crossref (an official Digital Object Identifier (DOI) Registration Agency of the International DOI Foundation, see www.crossref.org) and ROAD (a subset of the ISSN Register containing resources in OA with an ISSN number, see http://road.issn.org/). The concluded that “preliminary results suggest that identification of the OA state of publications denotes a difficult and currently unfulfilled task” (Akbaritabar & Stahlschmidt, 2019, p. 1).

Two other recent studies underlined not only the usefulness of DOAJ as data source on the access status of scientific journals, but also its limitations. In a study of publication strategies of Finnish researchers, Pölönen et al. (2019) found that, although publishing in DOAJ-indexed journals “is a good predictor of open availability of outputs”, they concluded that “our findings suggest that relying on external information sources, such as DOAJ, in the identification of open access publications may not result in complete picture of Gold OA publishing” (Pölönen et al., 2019, p.1785-1786). Björk (2019) noted in a second study on OA publishing in Nordic countries: “The results show that, although DOAJ provides very useful information about the OA journals published in the five Nordic countries, only 42% of the total number found in this study are included in the index” (Björk, 2019, p.233). Sivertsen et al. (2019), analysing the use and potential of Gold Open Access, underlined the changes in DOAJ, noting that “in March 2014, DOAJ launched a new and more stringent set of criteria for inclusion, leading to rejection of many journals that were previously included” (Sivertsen et al., 2019, p 1600).

In their study on converting to OA that will be discussed below, Matthias, Jahn & Laakso (2019) used another data source on OA, The Open Access Directory (OAD). OAD is a compendium of simple factual lists about open access (OA) to science and scholarship, maintained by the OA community at large since 2008 (OAD, n.d.). The authors qualify OAD “as the most extensive resource documenting such journal conversions is a wiki-page of OAD” (Matthias, Jahn & Laakso (2019, p. 2). Rimmert et al. (2017) present a database on Gold OA journals created at the University of Bielefeld (see https://pub.uni-bielefeld.de/record/2913654).

*Earlier studies on the adherence towards OA and the OA citation advantage*

One of the very first articles examining the effect of free online availability of research papers was published by Lawrence (2001). He found that “an average of 336% more citations to online computer science articles compared with offline articles published in the same venue” (Lawrence, 2001, p. 1). Ever since, numerous research papers have been published that measure an ‘OA Citation Advantage’, often abbreviated as OACA.

Craig et al., (2007) reviewed the literature on the subject published prior to 2007. They concluded that article OA status *alone* has little or no effect on citations, and underlined the influence of selection bias (authors make their better works OA) and an early view effect (articles posted as preprint show a more rapid increase of citation counts after their formal publication date). While Davis et al. (2008) found no citation advantage for OA articles in the first year after publication, Gargouri et al. (2010) measured both for self-selected or mandated OA a substantial OA citation advantage, due to a quality bias (OA advantage is larger for more frequently cited OA papers), rather than a selection bias.

A meta-analysis published by Swan (2010) reviewed the outcomes of 31 studies, and found in a great majority of cases a positive OA citation advantage. Two large studies, one by Solomon, Laakso & Björk (2013) - analysing SCImago.com data on OA journals indexed in Scopus -, and a second by Archambault et al. (2014) - based on Web of Science and studying 44 countries -, found positive OA citation advantages. Solomon, Laakso & Björk warned that the year a journal became OA included in DOAJ is self-reported and that “these dates are often inaccurate” (Solomon, Laakso & Björk, 2013, p.645). Their paper contains an Appendix with the assumptions of the criteria used in switching journals’.

A study by Sotudeh, Ghasempour & Yaghtin (2015) of OA journals published by Springer and Elsevier revealed positive citation advantages in *all* major research disciplines, while an analysis by Hua et al.

(2016) of dentistry journals “suggested no evidence that OA articles received significantly more citations than non-OA articles” (Hua et al., 2016, p. 48), and a study by McKiernan et al. (2016a) on 19 subject fields found an OA advantage in some fields, and a disadvantage in others. In a large study on OA, Piwowar et al. (2018) concluded that, “accounting for age and discipline, OA articles receive 18% more citations than average, an effect driven primarily by Green and Hybrid OA” (Piwowar et al, 2018, p. 1). De-Moya-Anegón, Guerrero-Bote and Herrán-Páez (2020) recently finished a study conducting a cross-national comparison and a cost-benefit analysis of Open Access models.

Analysing a sample of journals in economics and business studies, McCabe and Snyder (2015), identified weaknesses in the methodology used in the earlier studies of OA citation advantage studies. They claimed that the enormous effects found in previous studies were an artefact of their failure to control for article quality, disappearing once fixed effects are added as controls” McCabe and Snyder, 2015, p. 144). Lewis (2018) presents a comprehensive review of previous studies about the OA citation advantage (up to 2017). Holmberg et al. (2019) studies the relationship between a journal’s OA status and its altmetric activity as expressed in mentions in online platforms, calculating an OA *Altmetrics* Advantage.

An interesting, recent study by Copiello examines the effect upon the OA citation advantage of one specific factor: the degree of openness and inclusiveness of the database indexing a particular journal. The author concludes that the relationship is “controversial, also because of an uncertain boundary between OA and paywall articles (Copiello, 2019, p.995).

*Studies on OA flipping*

A special group of studies relates to OA “flipping” or “reverse flipping”. Solomon, Laakso & Björk (2016) introduced the term flipping to describe a journal’s conversion from non-OA to Gold-OA. They discussed the motivations behind the flips identified in the literature. They provided information about the literature on journal flips, especially on the motivations behind the flips, and incorporates expert interviews. In 2014, Busch published a blog on the effect of converting from non-OA to Gold-OA upon a journal’s impact factor (Busch, 2014). He reported an increase of the value of a journal’s impact factor, but a reduction of its publication output.

Matthias, Jahn & Laakso (2019) analysed the effect of both OA flipping and reverse flipping, i.e., conversion from Gold-OA to non-OA, upon journals’ citation impact. As indicated above, they used the Open Access Directory (OAD) as data source. They found during the time period between two years before and two years after the flip only slight changes in the impact of reverse-flipping journals. Momeni et al. (2019) analysed 171 journals indexed in the Web of Science switching from non-OA to Gold-OA, using the OAD Open Access Directory as OA data source, and the Web of Science as a citation source. They found that in most cases flipping had a positive effect on a journal's Impact Factor, but did not generate an obvious citation advantage for its articles. Also, they observed a decline in the number of published articles after flipping. Their overall conclusion holds that “flipping to open access can improve the performance of journals, despite decreasing the tendency of authors to submit their articles and no better citation advantages for articles" (Momeni et al., 2019, p. 1270).

*Research questions addressed in the current paper*

As is outlined in the review of the relevant literature above, many authors have conducted studies aimed to measure the so called *OA Citation Advantage* (sometimes abbreviated as OACA), and several papers have dealt specifically with flipping from non-OA to Gold-OA. Some of these also analysed whether there was a reduction in the publication output. If there is, one could speak of an *OA*

*Publication Disadvantage* (OAPD). The current study builds further upon these works, and adds the following features, that will be further explained in the data and methodology sections below.

- Many authors have underlined the difficulties in obtaining comprehensive, reliable data on the OA status of journals and individual articles. The current study compares two databases on OA status, and validates specific data elements from these sources.
- Analysing a trend in bibliometric indicators of flipping journals makes sense only when the outcomes are compared with those of non Gold-OA journals. Therefore, two well-defined control group were created, enabling paired statistical tests.
- A systematic analysis was conducted of changes in the distribution of publications and citations among income categories devised by the World Bank, and across geographical world regions.
- To distinguish short term and longer term effects, the length of the time period was varied during which flipping effects are monitored.
- A journal's national orientation was considered as an important factor influencing the impact of journals in general.
- Citation-indicators were calculated by SCImago Research Group based on Scopus, rather than on Web of Science.

The following research questions are addressed:

- How does the citation impact of journals switching from non-OA to Gold-OA develop after the flip year? Are there signs of an OA Citation *Advantage* and of an OA Publication *Disadvantage*?
- To which extent do the outcomes depend upon the OA data source used to identify OA flipping journals?

*Structure of the paper*

The next section provides information on OA and citation data sources that were combined to conduct the required analyses. It gives an outline of how the study sets of flipping journals were created. Next, the indicators calculated in the study are described. This section also presents the statistical tests that were applied, and describes how the control groups were created. The next section presents an analysis of DOAJ. Although several studies have underlined the limitations of this database as a directory of Gold Open Access journals, it is still functions as a standard in OA research, justifying a separate analysis in this paper. A detailed analysis of the statistical effects of flipping to OA, comparing two sets of flipping journals with two control groups, is presented in a subsequent section. In the final section conclusions are drawn, limitations of the current study are highlighted, and suggestions are made for further research.

## Data

*DOAJ-based approach*

The version of the DOAJ database used in the current study was downloaded from the DOAJ website on 15 December 2019, and contained on that day 14,052 journals. It was matched on the basis of journals' Print- and E-ISSN against a database created by SCImago Research Group that contained publication and citation data of journals indexed in Scopus during the time period 1996-2017 (see below). All about 20,000 Scopus journals were selected that were active in 2017, and that had published papers in each year between the first year for which they were indexed up until 2017. 14.8 per cent of DOAJ journals was linked with the selected Scopus journals. Almost 10 per cent of Scopus journals was found in DOAJ. An analysis presented below compares these DOAJ journals with non-DOAJ sources in Scopus. This is a secondary analysis, as the core analysis is based on the following two journal study sets.

The first study dataset is labelled as *DOAJ-based set (with verified OA switch year).* DOAJ contains a data field indicating the first calendar year a journal provided online Open Access content in DOAJ directory. In a first step, all 288 journals were selected from DOAJ for which this first year is between 2000 and 2013, and which according to Scopus/SCImago data had already published articles before that year. A working hypothesis was that this set contains candidate journals that switched from non-OA to Gold-OA, but that further validation was necessary. This hypothesis was tested in a verification round, in which emails were sent to publishers, editors-in-chief and journal contact information, asking them to indicate the year in which the journal switched to OA. Two reminders were sent. In the end, a response rate of 49 per cent was achieved (141 cases). 42 per cent of these 141 responses indicated an OA switch year that was different from the year indicated in DOAJ (i.e., the first calendar year the journal provided online Open Access content). It was decided to consider only DOAJ journals for which the response indicated a switch year. In this way, the analysis was entirely based upon *verified* data. But in view of the time period covered by the SCImago/Scopus data (1996-2017), only those journals were selected for which the switch year was between 2000 and 2013, and that had published at least one paper in the three years before switch year, and at least one paper in the three years after the switch. Thus, 22 journals from the set of 141 journals were dropped, so that the final DOAJ-based study set contained 119 journals.

*OAD-based approach*

A second study set was based on data from the Open Access Directory (OAD) mentioned in the introduction section, and labelled as *OAD-based Set*. The titles of all 296 journals indicated as converted or flipped from non-OA to Gold-OA were downloaded on 20 February 2020, as well as the year in which the conversion took place. 43 journals for which the switch year was missing were discarded. 200 journals had a switch year between 2000 and 2013, and were matched against the SCImago/Scopus journals on the basis of journal titles, taking into account the variations in titles. 135 journal titles appeared both in SCImago/Scopus and OAD. Of these, 35 of these had zero publications in the switch year or in the years before the switch, and were discarded, resulting in a final set of 100 journals. 73 of these are included in DOAJ. Interestingly, the overlap between this set and the *DOAJ-based set with verified OA switch year* was limited: 17 journals appeared in both. Therefore, it was decided to analyse both sets separately, and compare the outcomes.

*Bibliometric data from SCImago Research Group and Scopus.com*

The research described in the current paper used a special, intermediary dataset with publication and citation counts per journal derived from SCImago Journal Rank (SJR), a database with journal indicators based on data from Elsevier's Scopus. For all journals indexed in Scopus, and for the time period 1996-2017, data were available for each citing year on the number of publications made in the three preceding years, and on the number of citations in the particular year to these 1-3 year-old papers. Data were available for a journal as-a-whole, but also broken down by affiliation country of publishing or citing authors. From the Scopus Source Title List (Scopus, 2019) information was extracted on the publication languages of indexed journals. It was assumed that this information relates to the most recent year. To track changes in a journal's publication language over time, for each journal in the two sets of OA journals the total number of publications and the number of publications in English in the three years before, and in the three years after the OA switch year were extracted from Scopus.com.

## Indicators and statistical methods

*Indicators of journal impact*

The journal impact indicator used in the current study is a *relative* or *field-normalised citation* rate calculated as follows. Its basic scheme is similar to the Clarivate journal impact factor, but applies a different citation window: the average number of citations are counted in a particular year to 1-3 year-old documents published in the journal, rather than to 1-2 year-old documents. Document types included are articles, reviews, conference papers and short surveys This indicator is denoted as JIF. A field-normalized journal impact indicator is calculated by dividing a journal's JIF by the average JIF values for all journals in the same subject categories as those assigned to the journal, using a classification in Scopus of about 300 subject categories. If a journal was assigned to multiple subject categories, a weighted average JIF was calculated, the weights being determined to the number of subject category assignments to a journal.

*Indicator of national orientation (INO)*

A *first* indicator of national orientation of a journal relates to the authors publishing in the journal, and is defined as the percentage of papers (co-) authored by researchers from the country *publishing* the largest number of papers in the journal (symbol INO-P). For instance, if a journal has a INO-P value of 80 percent, this means that there is one country that accounts for 80 per cent of all papers published in that journal. It must be noted that a certain fraction of these papers can be expected to have authors from other countries as well, and thus reflect *international co-authorship*. A *second* indicator of national geographical orientation relates to the authors *citing* a particular journal (INO-C), and is defined as the percentage of citations given by researchers from the country contributing the largest number of citations to the journal's citation impact. For more details, the reader is referred to Moed et al. (2020).

*World Bank classification of countries based on income*

A classification made by the World Bank of countries by income level for the year 2019 was used to group countries into four classes: high income, upper-middle income, lower-middle income and low income, based on the Gross National Income (GNI) per capita (World Bank, 2019). The classification threshold in which the countries are classified is updated on an annual basis.

*Paired difference tests and outlier detection*

A paired difference test is used when comparing two sets of measurements to assess whether their population means differ. In the current study a paired T test is used. Although it assumes that the underlying variables are normally distributed, it is still valid in case of slight deviations from normality. Also, sample size increases its robustness despite the non-parametric nature of underlying data. To identify outliers, Tukey's criterion is used based on the interquartile range: if Q1 and Q3 are the lower and upper quartiles of the distribution of a variable's values, an outlier is any observation with a value outside the range $[Q1 - k(Q3 - Q1), Q3 + k(Q3 - Q1)]$, in which k is set to the value 2 in the current study.

*Trend analysis of journals switching to Gold-OA*

The analysis comparing indicators of a journal's publication output and citation impact in *pre-switch* years to those calculated for *post-switch* years focuses on the *relative* growth in the indicators rather than on *absolute* growth. Calculation of statistics on absolute growth introduces a strong "size" bias in favour of journals publishing large number of papers or having a high citation impact. Such a bias is would disadvantage journals that publish less or have a lower impact.

*Definition of control groups*

Defining a control or benchmark set is *not* a theoretically neutral activity, but does strongly determine the outcomes of any comparative analysis and their interpretation. The following considerations and factors were taken into account. They relate to the elimination of Gold or Hybrid OA journals from the control group, and the formation of *two* control groups based on different sets of controlling variables: a *standard* group, formed on the basis of a standard bibliometric criterion, and *tailor-made* group taking into account a journal's national orientation.

*i) Deleting OA journals from the universe of potential candidates for the control group.* If one aims to examine whether the *Gold* OA journals in the study set differ from *non Gold-OA* journals, one should eliminate from a control group all Gold-OA journals. Thus, in a first step all DOAJ journals were deleted from the universe of potential benchmarks from which a control group was formed. But if the objective is to compare the *Gold-OA* journals with *non-OA* journals in general, hybrid OA journals should be eliminated from a control group as well. In the current study, 1,280 journals in Scopus.com with more than 100 OA articles during 2014-2016 were identified. **SEE ENDNOTE 1**. According to SCImago data, 656 of these were not included in DOAJ; these were deleted from the benchmark universe as well. After a test identifying for DOAJ-based flipping journals potential benchmarks, and determining for these journals via Scopus.com the number of OA papers, an additional 698 journals were deleted for which the percentage of OA articles during 2014-2016 exceeded 10 %. The authors do *not* claim to have eliminated all hybrid OA journals, but at least a substantial part of it.

*ii) The formation of a standard control group.* A standard method to normalize bibliometric, citation-based indicators of a journal is to take into account its research discipline, and the publication year, and document type of the papers it published. In the current study no data were available on publication counts per document type. But data on covered research disciplines were readily available; the same is true for the key year in the analysis, namely the OA switch year. Therefore, a first control group was formed by selecting for each journal in a study set a benchmark journal covering the same research discipline as the study journal, and calculating for this benchmark indicators related to time periods before and after the same year as the OA switch year of the corresponding study journal. If for a particular study journal *multiple* benchmark journals were found, a benchmark journal was selected *at random*, applying a random number generator in the statistical package SAS. The thus formed control group will be denoted as the "*standard* control group".

iii) *Creating a tailor-made control group accounting for a journal's national orientation*. A study by Moed et al. (2020) found that a journal's national orientation is an important factor in determining its citation impact, even though the relationship between a journal's national orientation and its citation impact is found to be inverse U-shaped. Therefore, a second control group was formed, by selecting for each journal in the two study sets, and from the universe of candidate benchmark journals created in the formation of the *standard* control group, a benchmark journal having the same most productive author country, and showing the same decile value of the Index of National Orientation (INO, see definition in the section on indicators and statistical methods) as the study journal. Again, if for a given study journal *multiple* benchmark journals met the required criteria, the corresponding journal in the control group was selected *at random*.

*Additional comments on the definition of the control groups.*

An analysis presented above, based on a comparison between DOAJ and non-DOAJ journals, found evidence that the additional factors considered in the creation of the tailor-made control group, author country and national orientation, do have a significant statistical effect upon the outcome of the comparison, and thus provides additional evidence that taking these factors into account increases the information content of the analysis. Both the analysis of the DOAJ Directory as a whole, and also the study by Moed et al. (2020), found that *publication language* is an important factor as well. It was decided *not* to correct for publication language in the definition of the control group. The theoretical reason is that publication language is an editorial characteristic of the journal, that can be directly controlled by the publisher and/or editors. Correcting for publication language therefore can be conceived as giving "*bonus points*" to publishers/editors who intentionally limit the circulation and potential reading audience of their journals.

## Comparing DOAJ journals to non-DOAJ journals

This section presents characteristics of DOAJ journals compared to non-DOAJ journals, related to main discipline covered, national orientation, publication language and most frequently publishing countries. The journals in the two study OA sets analysed in the next section have similar characteristics. Therefore, the current section can be seen as an introduction to – and a justification of – the advanced analysis of OA flipping journals presented in the next section.

*Discipline, national orientation and publication language*

The analysis presented in Figures 1-4 below relates to Scopus journals included in DOAJ – denoted as DOAJ journals below – and to the publication time period 2014-2016. DOAJ Preference is defined as the difference between the actual and the expected number of DOAJ journals in Scopus (based on the independence model), expressed as a percentage relative to the expected number of DOAJ journals. For instance, the expected number of DOAJ journals in a discipline is calculated as the product of the overall fraction of DOAJ journals in Scopus (across all disciplines, relative to the total number of Scopus journals) and the absolute number of Scopus-covered DOAJ journals in that discipline. Figure 1 shows that in the set of DOAJ journals, those covering clinical medicine and natural sciences are overrepresented, and serials specializing in social sciences & humanities and engineering underrepresented, while Figure 2 makes clear that DOAJ journals tend to have a stronger national orientation in term of the geographical location of publishing authors compared to their non-DOAJ counterparts.

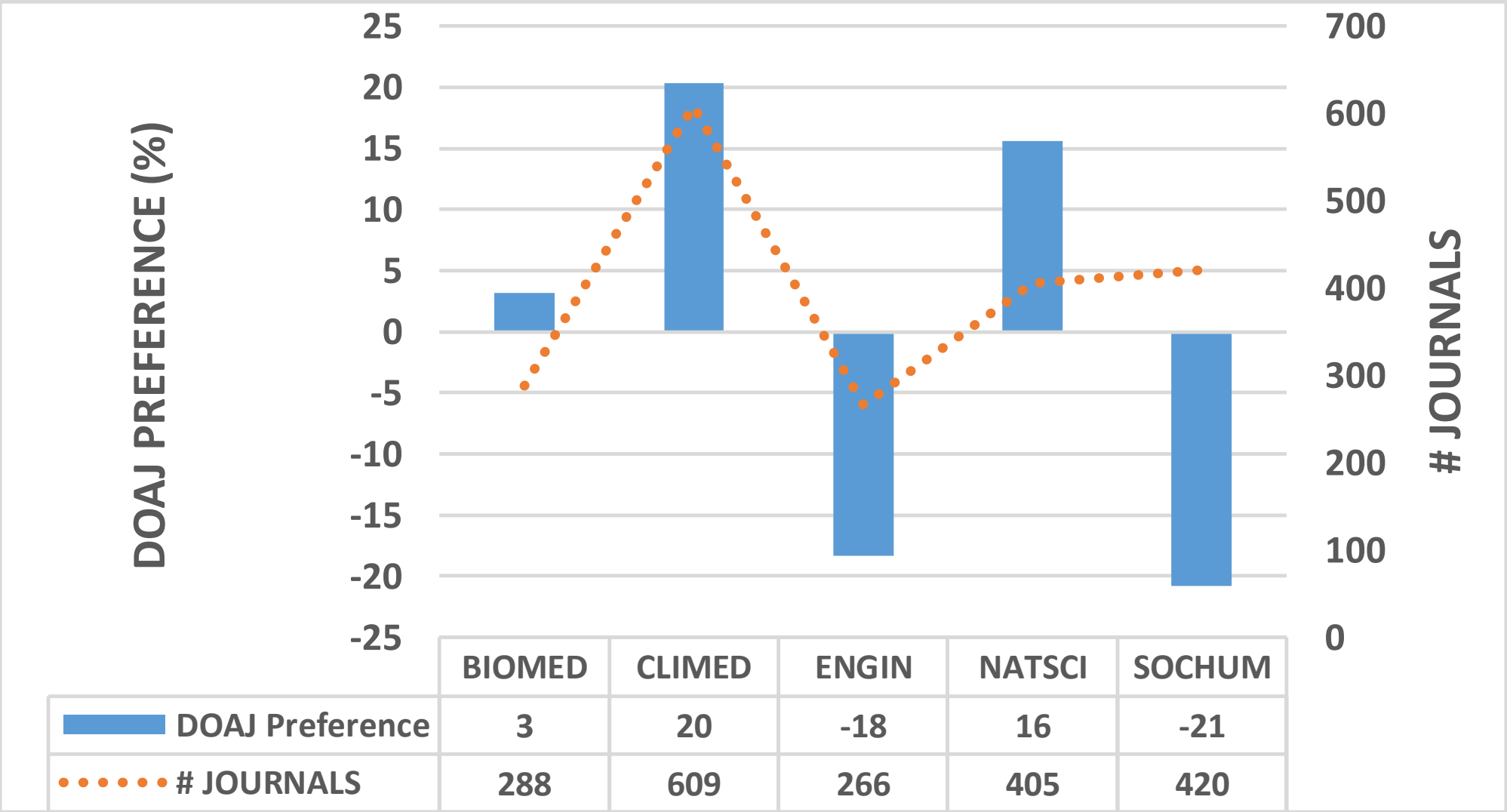

Figure 1: DOAJ Preference per main discipline. DOAJ Preference is defined as the difference between the actual and the expected number of DOAJ journals indexed in Scopus, relative to the expected number of DOAJ journals, and expressed as a percentage.

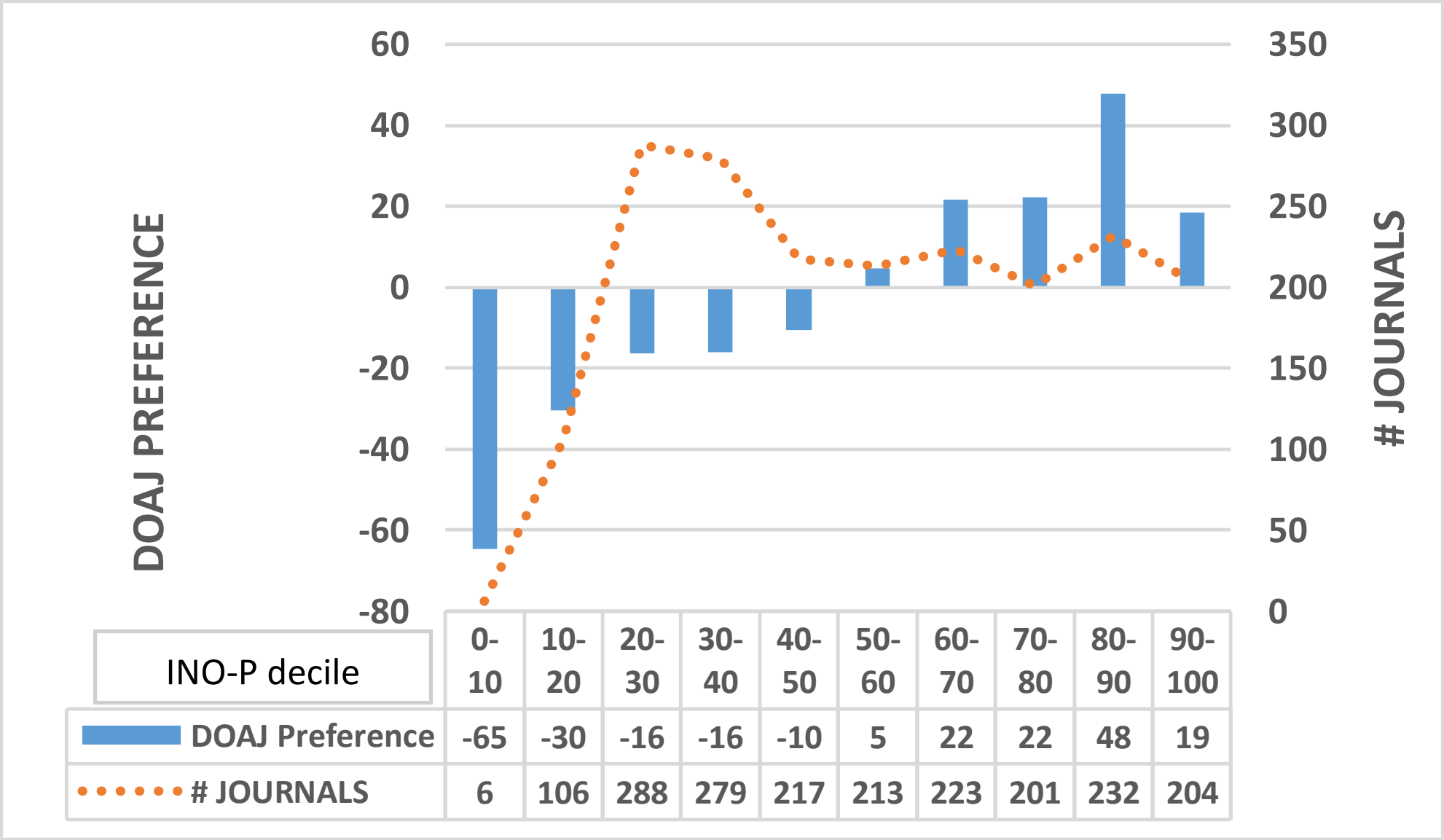

Figure 2: DOAJ Preference versus national orientation (INO-P decile).

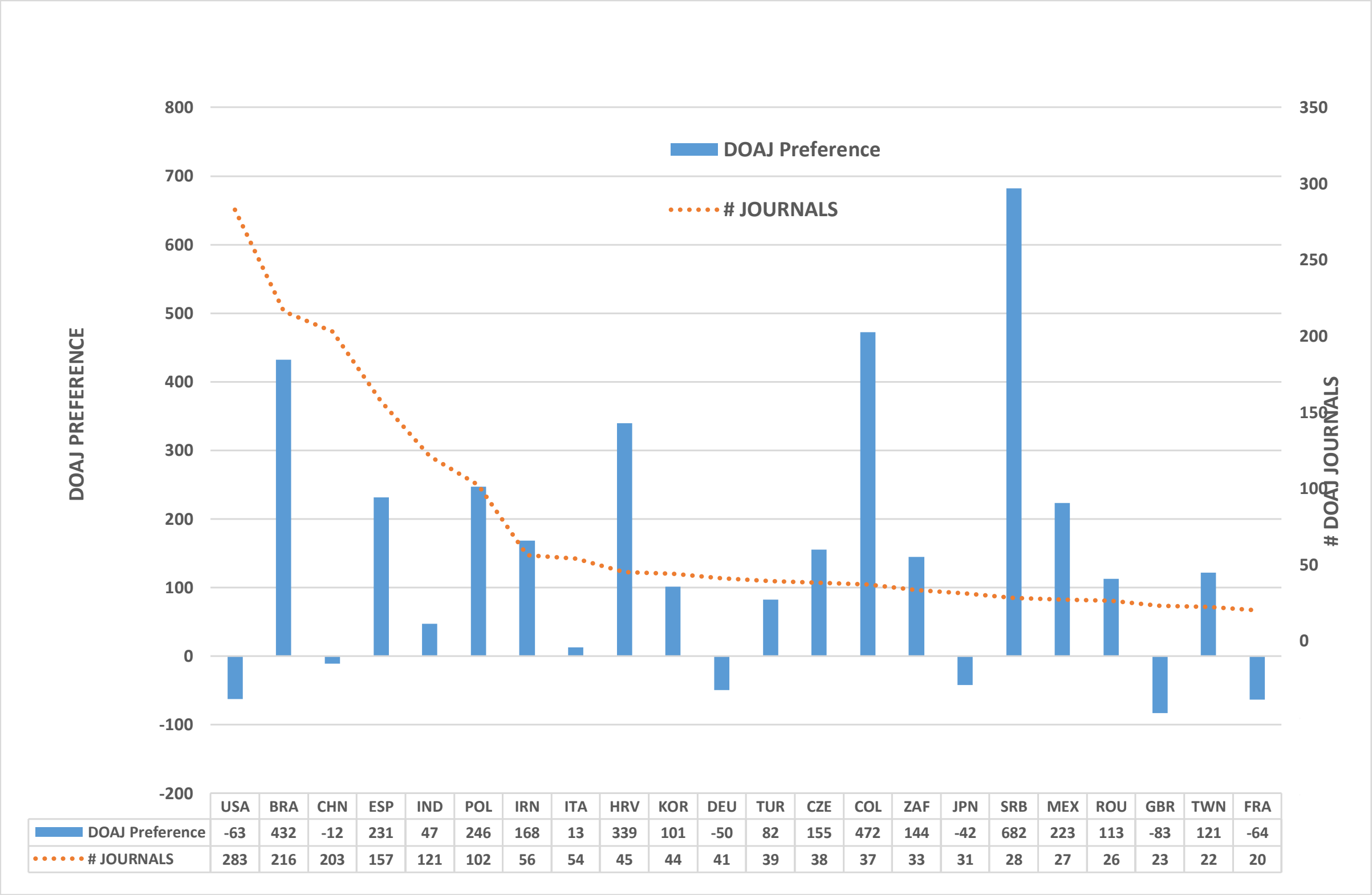

Figure 3. DOAJ preference by publishing author country for countries with 20 or more DOAJ journals in Scopus.

Next, Figure 3 shows for each country with 20 or more DOAJ journals in Scopus the number of DOAJ journals in which authors from a given country published the largest number of papers. For instance,

Brazil is the most productive author affiliation country in 216 DOAJ journals, and is highly overrepresented in the set of DOAJ journals compared to their appearance as most productive country in non-DOAJ journals. According to Leta et al. (2019), Brazil's outstanding position is due to initiatives at a national level, especially "the foundation and expansion of the Scientific Electronic Library Online (SciELO) during the last two decades, as well as the introduction of the Electronic Journaling System and the launching of the Brazilian Manifesto to Support Open Access to Scientific Information in 2005" (Leta et al., 2019, p. 1760). Other highly overrepresented countries (with DOAJ Preference above 200%) are Serbia, Colombia, Croatia, Poland, Spain and Mexico. The following countries are underrepresented (with DOAJ Preference below 0%): Great Britain, France, USA, Germany, Japan and China. Finally, Figure 4 analyses information on a journal's publication language from the Scopus Source List (Scopus, 2019) and reveals that journals publishing merely in non-English languages, or multi-lingual journals including English, are overrepresented in DOAJ.

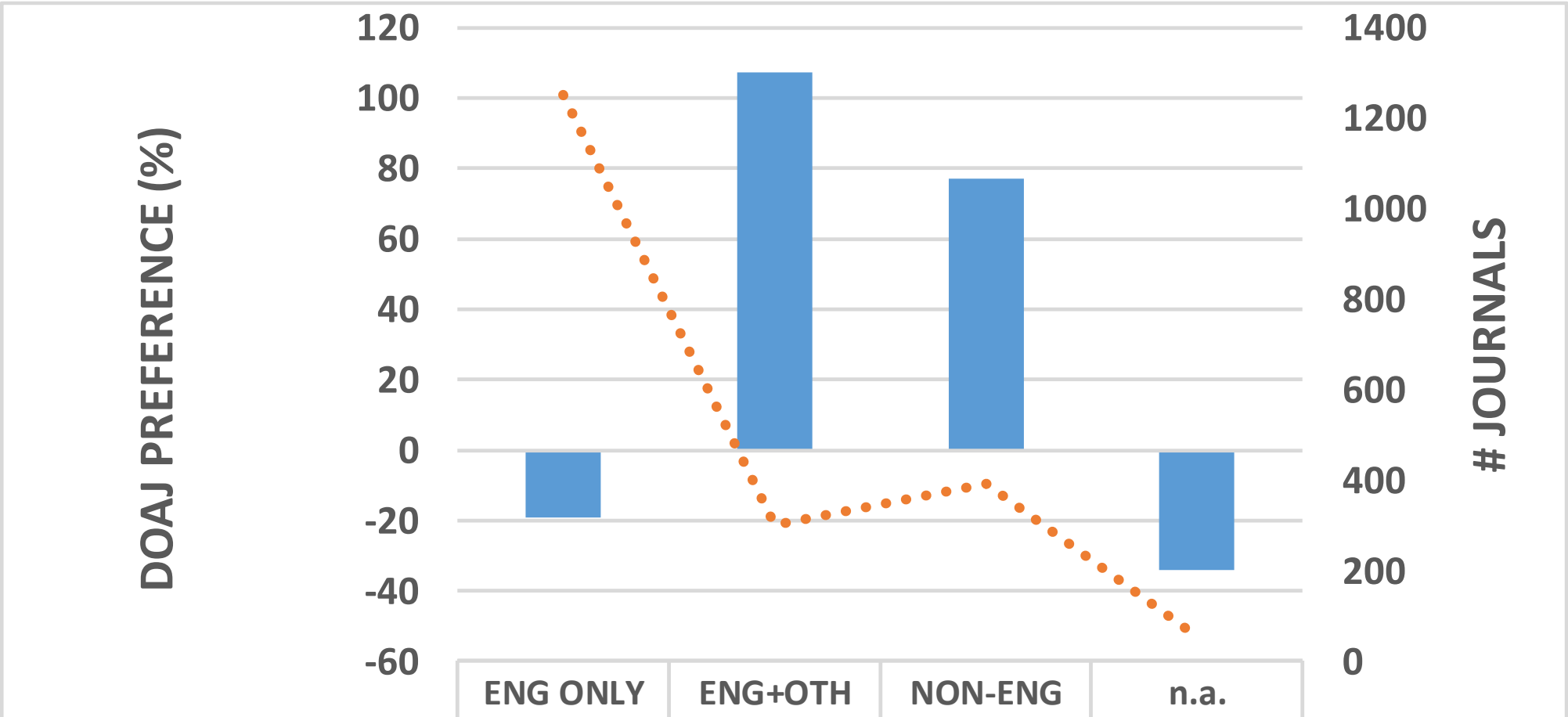

| | ENG ONLY | ENG+OTH | NON-ENG | n.a. |
|---|---|---|---|---|
| DOAJ Preference | -19.2 | 107.2 | 77.2 | -34.2 |
| # JOURNALS | 1252 | 300 | 392 | 56 |

Figure 4. DOAJ Preference by publication language. ENG ONLY: Publication language is English for all content. ENG+OTH: Published both in English and in non-English languages. NON-ENG: Publishes merely in non-English languages. N.a.: No information available.

*Analysis of affiliation countries of publishing and citing authors*

How do the distributions of articles in DOAJ journals across *income classes* and *geographical regions* of the affiliation countries of publishing authors differ from those distributions in non-DOAJ journals? The results are presented in Figure 5.

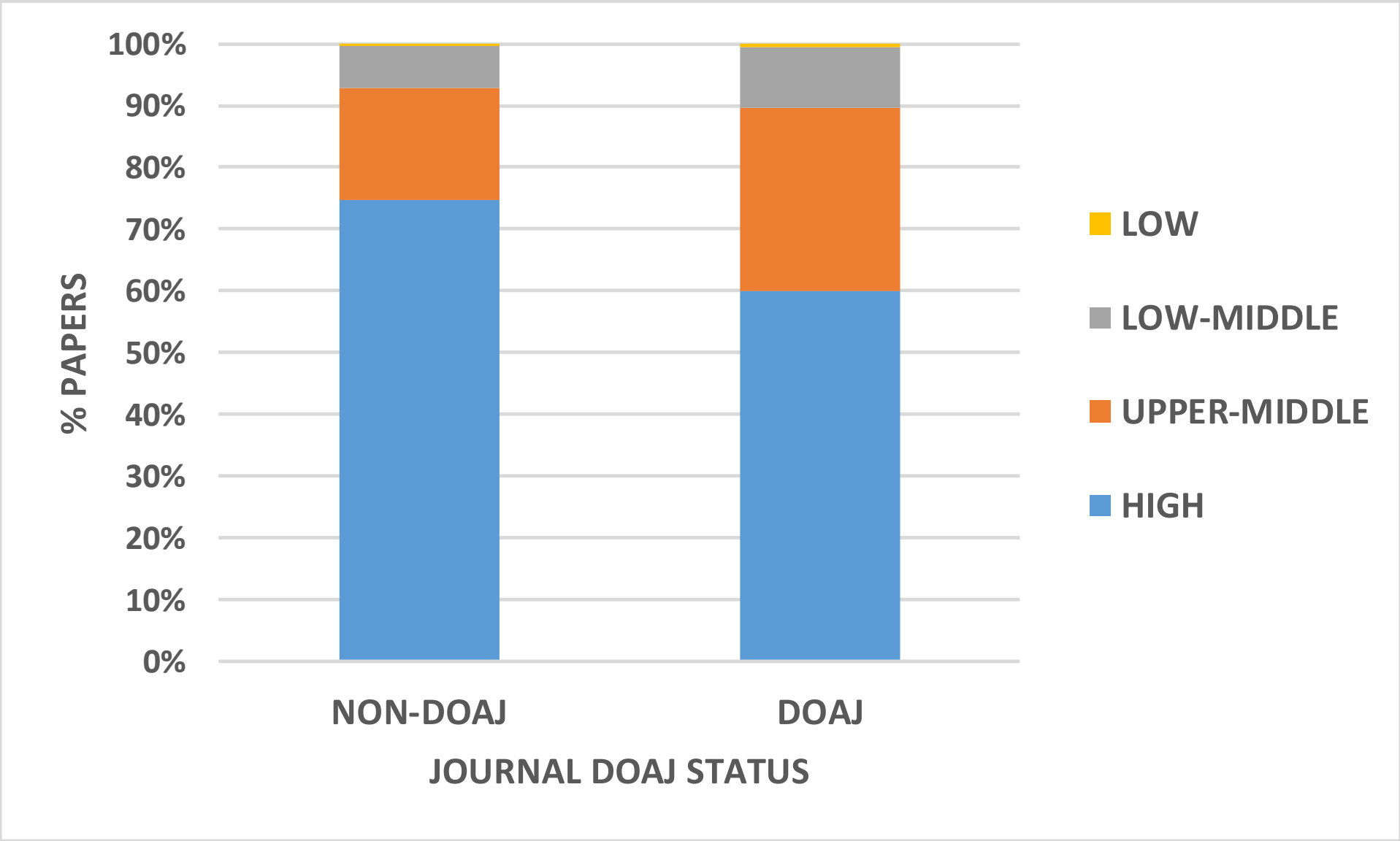

Figure 5: Distribution of articles in DOAJ journals across income classes of the affiliation countries of publishing authors. Total number of articles published during 2014-2016 (including double counts due to international co-authorship) in DOAJ journals indexed in Scopus: 11,452; in non-DOAJ journals indexed in Scopus: 99,772. Data on income class obtained from the World Bank (2019).

The percentage of articles from low-middle and low income countries is in DOAJ slightly higher than it is in the complementary set of non-DOAJ journals: 10.3 against 7.2 per cent. The share of papers in DOAJ from upper-middle countries is substantially higher: 30 against 18 per cent.

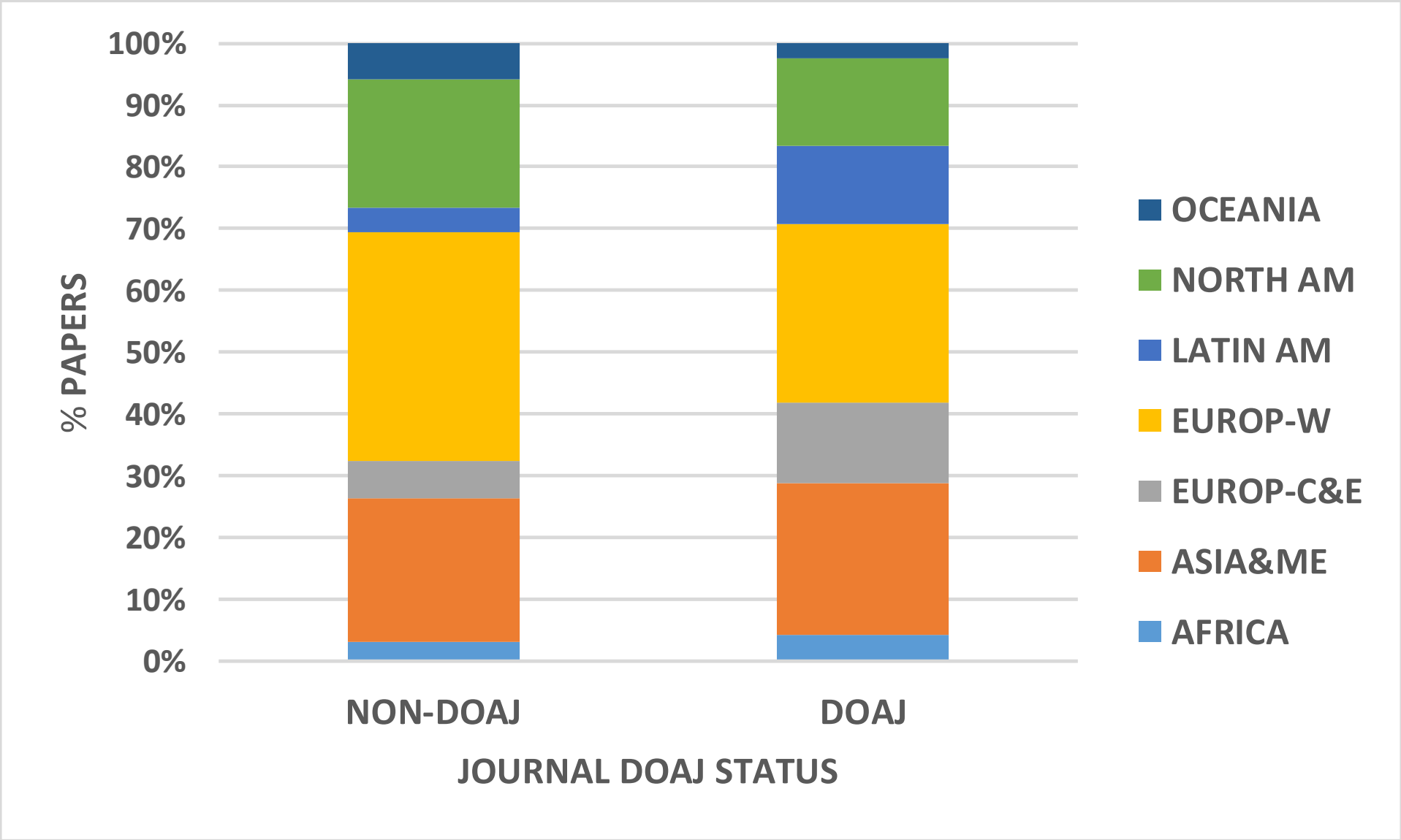

Figure 6: Distributions of articles in DOAJ journals across *world regions in the affiliation countries of publishing authors.*

A striking feature in Figure 6 is the relatively large share in DOAJ journals of papers from Central & Eastern European, and from Latin American countries, at the expense of the article share from Western Europe and Northern America, respectively. By contrast, the share of papers from Oceania reveals a large share in non-DOAJ instead of DOAJ.

*Comparing the mean relative citation impact of DOAJ and non-DOAJ journals*

The mean relative citation impact in the (citing) year 2017 of DOAJ journals was compared with that non-DOAJ journals, applying a T-test. A comparison of the means of the two samples shows that *non-DOAJ* journals have on average a *highe*r citation impact than their DOAJ counterparts (1.01 versus 0.77). Since the underlying distributions deviate strongly from normality, a T-test was conducted based on the logarithmic values of the impact scores were calculated. Moreover, the Satterthwaite method was used as the assumption that the two populations have equal variances seems unreasonable. For non-DOAJ and DOAJ journals the values -0.52 and -0.67 were obtained, respectively, a difference of 0.15 which according to the Satterthwaite T-test significant at p<0.001.

*Regression analysis*

A dummy variable was created with value 1 if a journal is included in DOAJ and 0 otherwise. In the same manner, dummy variables were calculated for having English as sole publication language, and for the 13 most important author countries. The dependent variable in the regression analysis is the relative (field-normalized) citation impact in 2017 of a journal's papers published during 2014-2016.

Taking into account not merely whether or not a journal is included in DOAJ, but also its publication language, national orientation, starting year of being indexed in Scopus, and its most productive author country, and defining for all categorical variables dummy variables, a picture emerges that is different from that generated by the paired T-test presented above. Tale 1 shows that the effect of being included in DOAJ upon a journal's citation impact is positive and significant (regression parameter b=0.140). For most countries the effect is positive and significant, but the size of the effect as reflected in the parameter estimates differs among author affiliation countries. It is especially striking in USA (b=0.435), China (b=0.302) and Great Britain (b=0.253). For India, Poland and Russia it is negative, while for France it is positive but not significant. It should be noted that the number of DOAJ journals in which Russia and France are the most productive author country is rather low (20 and 19 journals, respectively). Publishing solely in English has the largest effect, positive and highly statistically significant, in full agreement with earlier studies on the effect of publication language.

The effect of national orientation is small though significant. It must be noted that Moed et al. (2020) found that the statistical relationship between citation impact and Index of National Orientation is *inverse-U shaped*. This means that linear regression is not an appropriate method to examine the statistical effect of this factor. The negative effect of a journal's first publication year indexed in Scopus indicates that journals recently indexed in Scopus tend to have a somewhat lower impact than older journals. Although this result is difficult to interpret, it does provide a justification for taking into account the OA switch year in the definition of the control groups analysed in the next section.

Table 1. Linear regression with a journal's citation impact as dependent variable

| *Variable* | | *DF* | *Parameter Estimate* | *Standard Error* | *t Value* | *Pr > \|t\|* |
|---|---|---|---|---|---|---|
| Intercept | | 1 | 36.604 | 0.741 | 49.39 | <.0001 |
| Included in DOAJ (dummy var) | | 1 | 0.140 | 0.006 | 21.62 | <.0001 |
| English Only (dummy var) | | 1 | 0.332 | 0.006 | 58.85 | <.0001 |
| National Orientation (INO) | | 1 | -0.004 | 0.000 | -48.63 | <.0001 |
| Journal starting year in Scopus | | 1 | -0.018 | 0.000 | -48.71 | <.0001 |
| Most productive author country (dummy var) | USA | 1 | 0.435 | 0.006 | 78 | <.0001 |
| | Brazil | 1 | 0.090 | 0.013 | 6.68 | <.0001 |
| | China | 1 | 0.302 | 0.007 | 42.68 | <.0001 |
| | Spain | 1 | 0.142 | 0.013 | 11.34 | <.0001 |
| | India | 1 | -0.114 | 0.010 | -11.73 | <.0001 |
| | Poland | 1 | -0.031 | 0.016 | -1.99 | 0.0469 |
| | Iran | 1 | 0.084 | 0.018 | 4.76 | <.0001 |
| | Italy | 1 | 0.139 | 0.012 | 11.12 | <.0001 |
| | Korea | 1 | 0.209 | 0.018 | 11.75 | <.0001 |
| | Germany | 1 | 0.124 | 0.010 | 12.25 | <.0001 |
| | Great Britain | 1 | 0.253 | 0.008 | 30.21 | <.0001 |
| | France | 1 | 0.015 | 0.012 | 1.27 | 0.2035 |
| | Russia | 1 | -0.097 | 0.014 | -6.91 | <.0001 |

## Analysis of journals switching to Gold-OA

This section presents a series of analyses of journals flipping to OA:

- Two *study sets* are being analysed, denoted as "*DOAJ-based set (with verified OA switch years)*" (n=119) and the "*OAD-based*" set (n=100). The reader is referred to the section on data for details.
- Each study set is compared with two *control groups*, a "*standard* control group" controlling for *discipline and age*, and a "tailored control group", controlling in addition for *national orientation*. The section on indicators and statistical methods presents the details.
- Next, two *performance aspects* are assessed: an *OA Citation Advantage*, using citation rates, and a *OA Publication Disadvantage*, based on publication counts. See the introduction section for a review of earlier studies.
- Moreover, the current section also compares the geographical distribution of authors publishing in the journals, in terms of an affiliation country's income status and its geographical world region.
- *Paired T-tests* were applied, comparing a study set with a control group, and the state of a study journal before its OA switch year with that of the same set after the switch.
- In the assessment two *time windows* are applied: a *short term* window, measuring possible effects during the first three years after a journal's OA switch year, and a *longer term* one, measuring affects during the fourth until seventh year after the flip.

It must be noted that the trend analysis is *not* affected by the phenomenon that journals switching to OA in a particular year may make their entire backlog OA as well. The publication and citation counts for the pre-switch period relate to publication and citation years *before* the switch year. Making the backlog OA in the switch year can only affect the citation counts of backlog papers *after* the switch year, *not* those received before the switch year.

The structure of this section is as follows. First, Table 2 presents basic statistics on the number of journals and the growth rates of the citation and publication-based indicators for the two study sets and the 4 control sets separately. Next, Table 3 presents the outcomes of a paired T-test, in which each study journal is compared on a one-to-one basis with its corresponding benchmark journal from one of the control groups. Finally, graphs show the geographical distributions of publishing authors in terms of the income class and world region of their affiliation countries.

*Basic statistics per journal set*

Table 2. Growth rates for study sets and control groups for all 4 indicators

| Journal Set | Citation Impact | | # Publications | |
|---|---|---|---|---|
| | Short term | Longer term | Short term | Longer term |

| | # Journals | Av. Growth rate | # Journals | Av. Growth rate | # Journals | Av Growth rate | # Journals | Av Growth rate |
|---|---|---|---|---|---|---|---|---|
| DOAJ-based Study Set | 114 | 0.505 | 55 | 0.626 | 113 | 0.174 | 57 | 0.414 |
| Standard Control Group | 111 | 0.107 | 66 | 0.239 | 110 | 0.149 | 65 | 0.219 |
| Tailored Control Group | 110 | 0.145 | 61 | 0.198 | 116 | 0.184 | 64 | 0.293 |
| | | | | | | | | |
| OAD-based study set | 94 | 0.349 | 61 | 0.570 | 95 | 0.111 | 57 | 0.152 |
| Standard Control Group | 97 | 0.149 | 64 | 0.258 | 97 | 0.114 | 63 | 0.161 |
| Tailored Control Group | 88 | 0.045 | 55 | 0.018 | 97 | 0.162 | 61 | 0.165 |

Legend to Table 2. Time windows: comparing the score in $4^{th}$ year (short term) or in the $7^{th}$ year (longer term) after switch year with the score in the year before switch year. Number of journals: outliers and journals with missing values on one or more variables are not included. Given the time period covered by the bibliometric data (1999-2017), the calculations for longer term indicators only include study journals switching before 2011. Growth rate: the mean value over all journals in a set of the logarithm of the ratio of a journal's score in the post-switch year and that in the pre-switch year. Standard Control Group: contains benchmark journals in the same discipline and data for the same years as those for the corresponding OA journal. Tailored control group: is formed on the basis of a study's journal's national orientation. Standard deviations exceed the mean values in all cases, often with a factor of 2.

Table 2 allows for the following observations.

- For both OA study sets and for both citation windows, OA journals tend to show on average a higher growth rate in their citation impact than the journals have in each control group.
- As regards the trend in the number of publications, differences exist between the two sets of OA journals: the average growth rate of publications is for journals in the OAD study set lower than that in this study set's control groups, while for the DOAJ study set the short term trend is lower, and the longer term trend higher than those of the DOAJ control groups.
- Both for citation and publication-based indicators, the average growth rate of the journals in de DOAJ study set is larger than that for the OAD study set.

*Results of the paired T-test*

Table 3: Comparison between study sets and control groups according to the increase in their citation impact and publication output

| Study Set | Variable | Standard Control Group | | Tailored Control Group | |
|---|---|---|---|---|---|
| | | # Journals | Mean Difference (study–control group | # Journals | Mean Difference (study–control group |
| DOAJ-based Study Set | Cit impact in 4th yr after OA Switch Yr | 111 | 0.382** | 109 | 0.391** |
| | Cit impact in 7th yr after OA Switch Yr | 54 | 0.336* | 49 | 0.366** |
| | # Articles in first 3 yrs after OA Switch yr | 116 | 0.039 | 116 | 0.025 |
| | # Articles in 4-6th yr after OA Switch yr | 55 | 0.157 | 56 | 0.129 |
| | Cit impact in 4th yr after OA Switch Yr | 95 | 0.210* | 86 | 0.308** |
| OAD-based Study Set | Cit impact in 7th yr after OA Switch Yr | 63 | 0.411** | 58 | 0.418** |
| | # Articles in first 3 yrs after OA Switch yr | 94 | -0.083 | 98 | -0.037 |
| | # Articles in 4-6th yr after OA Switch yr | 60 | -0.021 | 60 | 0.007 |

Legend to Table 3. The rate of increase was defined as the logarithmic value of the ratio of the post-switch and pre-switch values, applying a short term and a longer term time window. Number of journals: outliers and journals with missing values not included. *: significant at p<0.05 (but not at p<0.01). **: significant at p<0.01.

While Table 2 compares a journal set "with itself", Table 3 compares the two study sets with each of the two control groups on the basis of the increase in citation impact and publication output values after flipping to OA, relative to those before the switch. The following observations can be made.

- Both the DOAJ-based and the OAD-based study sets show the following pattern: journals switching to OA tend to significantly increase after the switch their citation impact more rapidly than journals in all control groups do.
- As regards the trend in publication counts, the paired T test presented in Table 3 shows that the differences in increase rates between study and control set are *not* statistically significant.
- Compared to the trends in the control groups, the two study sets show only minor differences between short term and longer term trends.

*Analysis of affiliation countries of publishing and citing authors*

Figure 7 relates to the DOAJ-based study set and its two control groups. It analyses for all three sets the distribution of publishing and citing authors among the income class of their affiliation countries, comparing the situation before the journals' switch to OA with that after the switch. Figure 8 displays for each income class the absolute difference between the post-switch and a pre-switch share of authors. The following observations can be made.

- For all three sets the share of authors from high income countries declines after the switch, while that for upper-middle incomes increases. This is true both for publishing and for citing authors. Figure 8 reveals that the DOAJ study set and the two control sets show more or less the same pattern.
- In the DOAJ-based study set the share of authors from low-middle income countries remains approximately constant, while for the benchmark sets it increases after the switch compared to the pre-switch situation.
- The share of publishing or citing authors from low income countries is extremely low in all three sets.

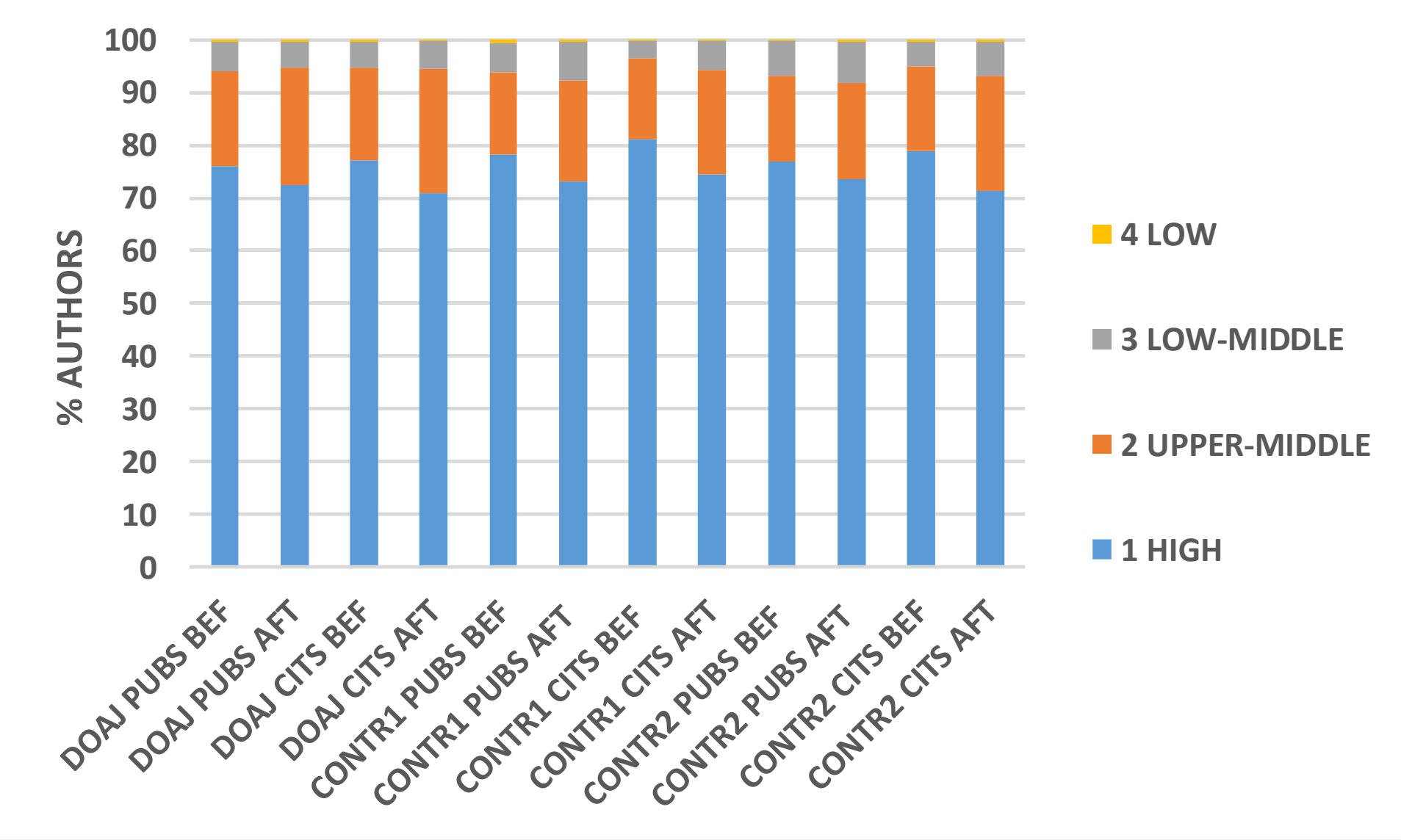

Figure 7. Distribution of publishing and citing authors among the income classes of their affiliation countries. DOAJ refers to the DOA-based Study Set, CONTR1 to the standard control group, and CONTR2 to the tailed control group. PUBS and CITS ref to the distribution related to publishing and citing authors, respectively. BEF and AFT mean 'Before' and 'After OA Switch Year', respectively.

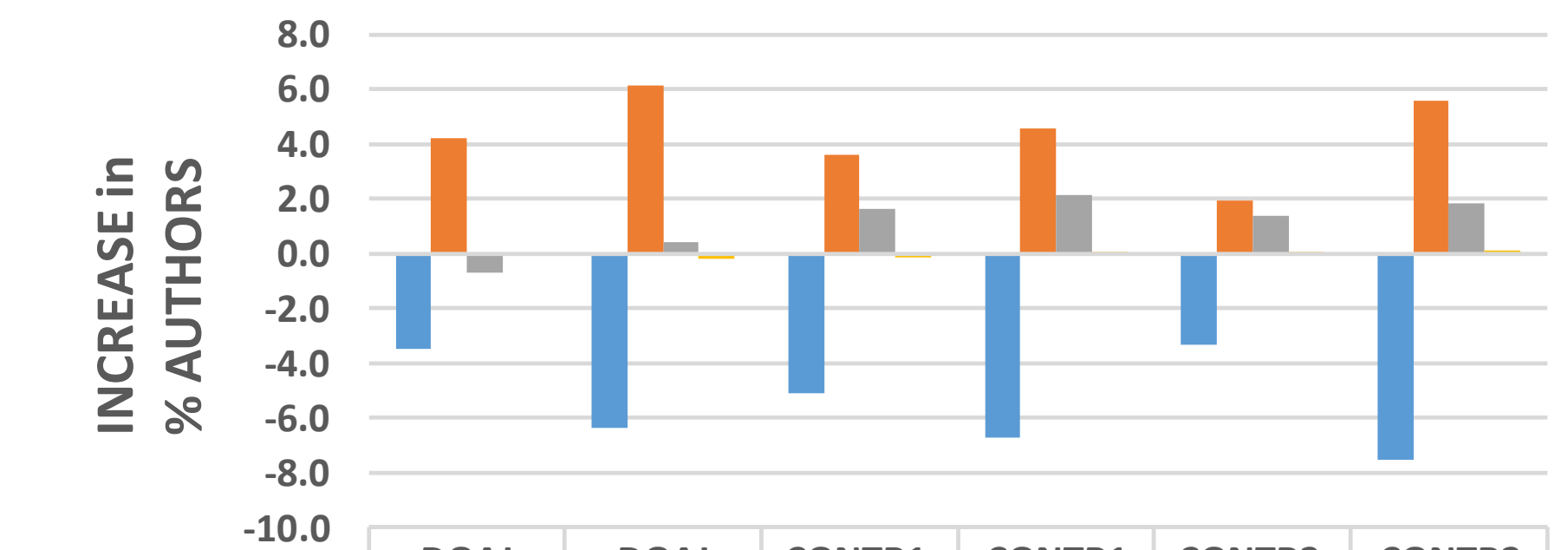

| | DOAJ PUBS AFT-BEF | DOAJ CITS AFT-BEF | CONTR1 PUBS AFT-BEF | CONTR1 CITS AFT-BEF | CONTR2 PUBS AFT-BEF | CONTR2 CITS AFT-BEF |
|---|---|---|---|---|---|---|
| 1 HIGH | -3.5 | -6.4 | -5.1 | -6.7 | -3.3 | -7.5 |
| 2 UPPER-MIDDLE | 4.2 | 6.1 | 3.6 | 4.6 | 1.9 | 5.6 |
| 3 LOW-MIDDLE | -0.7 | 0.4 | 1.6 | 2.1 | 1.4 | 1.8 |
| 4 LOW | 0.0 | -0.2 | -0.1 | 0.0 | 0.0 | 0.1 |

Figure 8: Increase in the share of publishing and citing authors in the various income classes. The increase is defined as the absolute difference between the post-switch share and the pre-switch share. DOAJ refers to the DOAJ-based Study Set, CONTR1 to the standard control group, and CONTR2 to the tailed control group. PUBS and CITS ref to the distribution related to publishing and citing authors, respectively.

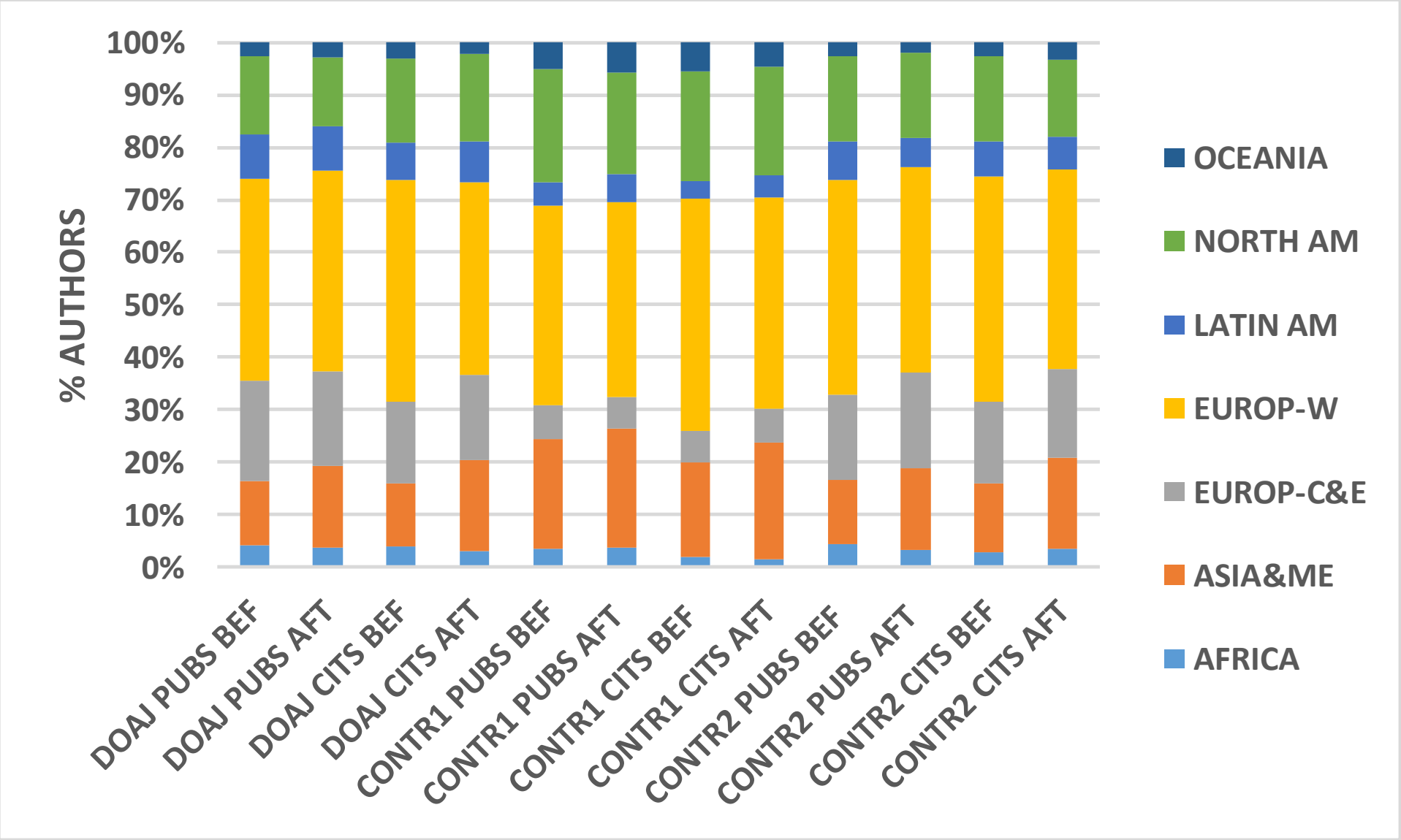

Figure 9. Distribution of publishing and citing authors among the geographical regions of their affiliation countries. DOAJ refers to the DOAJ-based Study Set, CONTR1 to the standard control group, and CONTR2 to the tailed control group. PUBS and CITS ref to the distribution related to publishing and citing authors, respectively. BEF and AFT mean 'Before' and 'After OA Switch Year', respectively.

Figure 9 is similar to Figure 7, but analyses the distribution of authors among geographical regions. It allows for the following observations.

- The differences between the pre- and the post switch distributions tend to be smaller than those related to income class displayed in Figure 7. The share of authors from Asia & Middle East and Latin America tends to increase, both in the study set and in the two control groups, while that for of Western-European and North-American authors declines. An analysis similar to that presented in Figure 8 reveals that study set and control groups show about the same pattern.
- Striking differences exist between the standard and the tailored *control group*. The shares of publishing and citing authors from Asia & Middle East, Oceania and from North America tend to be higher in the standard control group than they are in the tailored group, and the share of Central and Eastern European authors much lower. These differences are due to differences in the definition of the two control groups. The standard group, containing journals selected merely on the basis of a study journal's discipline (and switch year), is a reflection of the distribution of journals among countries in *the Scopus database*, while the tailored group, as it takes into account a study journal's national orientation, is therefore more similar to the geographical distribution of authors *in the study set*.

## Discussion and conclusions

*General conclusions*

*Inconsistencies in databases on OA*.

The work described in the current paper has confirmed the data problems encountered by other researchers studying the OA phenomenon. The two major ones were as follows. In the sample studied, the data in a DOAJ in a field labelled "the first calendar year a journal provided online Open Access content" differed in about 40 per cent of cases from the year indicated by the publishers in a verification round. These results should encourage these directories to offer reliable information upon their flipping year. Next, the list of flipping journals in the Open Access Database (OAD) contains journals that could *not* be found in DOAJ.

*Gold OA journals have special characteristics*.

The case study on the Directory of Open Access Journals (DOAJ) showed that journals in DOAJ covering clinical medicine and natural sciences are overrepresented, and serials specializing in social sciences & humanities and engineering underrepresented; DOAJ sources tend to have a stronger national orientation, and use more often non-English as a publication language than non-DOAJ journals, and are more oriented towards Latin America and Central- and Eastern Europe, and less strongly towards North America, Western Europe and Oceania.

*The definition of control groups.*

The study aimed to compare journals switching to Gold OA with non Gold-OA journals, and, as accurately as possible, with non-OA journals. Therefore, hybrid journals with large numbers of article-based OA content were deleted from the universe from which control journals were drawn. It was acknowledged that defining a control group is *not* a theoretically neutral act. In view of the above mentioned characteristics of OA journals, it was decided to define two control groups: a *standard* control group, correcting only for discipline and OA switch year of the OA journal, and a second set controlling also for a journal's national orientation. In this way, two study sets were compared with two benchmark sets.

If the outcomes of the current study are indeed affected by the inclusion of hybrid is journals in the control groups, their full elimination could make the observed differences between study journals and control groups even larger. This would be the case if hybrid journals increase their impact by making a part of their papers OA, especially if this portion increased over time. To examine this issue further, a longitudinal study of hybrid OA journals, and a direct comparison between Gold OA and other OA forms would be necessary.

*Conclusions on OA Citation Advantage.*

Both sets of journals switching to Gold-OA tend to increase their citation impact more rapidly than each of their two control groups do. This outcome provides robust evidence, independent of the OA data source used, that flipping journals tend to have an OA Citation Advantage compared to *non Gold-OA* journals, and that this advantage is already visible a few years after the switch. It corroborates the conclusions from a series of previous studies on OA flipping, mentioned in the introduction section. Whether this citation advantage also occurs in relation to *non-OA journals in general, excluding hybrid OA serials*, remains to be assessed in a follow-up study, that is further discussed below. All that can be stated in the current study is that voluminous hybrid OA journals were at least partly eliminated as candidates from the control group.

*Conclusions on OA Publication disadvantage.*

The paired T-test found no statistically significant evidence that journals switching to OA increase their publication output faster than non Gold-OA journals do. Therefore, there is no solid evidence for an OA Publication *Advantage*, nor for an OA Publication *Disadvantage*. This outcome is not in agreement with earlier studies reporting a decline of the publication counts of flipping journals after the switch. But although the differences in outcomes between the two sets compared to their benchmarks are not statistically significant according to the paired T-test, some traces of a lower growth of journals in the OAD study set compared to its control groups were found, the lack of statistical significance of which could perhaps be ascribed to the relatively small sample sizes. The fact that these traces are found in the OAD but *not* in de DOAJ-based study set underlines the need to further examine and improve the data quality and consistency of data sources on Open Access.

*Analysis of affiliation countries of publishing and citing authors*

Analysing the affiliation countries of authors publishing in journals in a particular set or citing them, and comparing the situation after the switch to OA with that before the switch, it was found that the share of authors from high income countries declines, while that for upper-middle incomes increases. But there are *no apparent* differences between de DOAJ study set and the two control groups in this respect. At the same time, there is an increase in the share of authors from Asia & Middle East and Latin America, at the expense of that of authors from Northern America and Western Europe. Also in this analysis, no differences were found between the OA study journals and the control groups. A plausible explanation holds that these shifts in relative importance of regions and countries primarily reflect changes in the *database coverage* rather than in access modality.

*The role of low and middle-low income countries*

According to the analysis of the distribution of publishing or citing authors among income class of their affiliation countries, the share of authors from middle-low and especially from low income countries is very low. In the two study sets there is no increase in the share of publishing or citing authors from these countries. The following comments provide a context for this outcome. Firstly, to become visible in this type of analysis, countries should be able to substantially increase their research efforts. Possibly low and low-middle income countries have not yet been able to do so. Secondly, major publishers have special arrangements with these countries, offering all their content at a low cost, if not for free. OA may therefore not be a primary concern to authors in these countries.

*Limitations*

It should be underlined that the analysis covers only a small part of DOAJ: only 15 per cent of DOAJ journals are in Scopus. Also one must consider the limitations of using the Scopus database as a data source, such as possible language biases or field biases. It cannot be taken for granted that results based on Web of Science would be similar to those based on Scopus presented in the current paper. Moreover, the role of publication language was not systematically investigated in the current study. As already mentioned in the text, although a substantial number of hybrid OA journals was eliminated from the control groups, the possibility that such journals have some effect upon the outcomes cannot be excluded. Therefore, the conclusions drawn is the study are based on a comparison between journals switching to Gold-OA with sources that are not Gold-OA, rather than with journals that have no OA at all.

*Suggestions for further research*

The current study is purely statistical, as it examines data samples, and identifies statistical tendencies in the data. To obtain more insight into the flipping process, especially into the contextual factors that influence its success, it is recommended to conduct also descriptive, qualitative case studies of individual journals switching to OA, in which not only switching from non-OA to OA, but also reversed switching from OA to non-OA could be considered. Potentially relevant factors such as the level of article processing charges (APC), types of publishers, disciplinary profile of countries' output or the duration of the publication process should be studied. Special attention could be given to possible interactions between influential factors that could be explored with decision tree techniques. If a journal switches to Gold OA, the change in access modality may at the same time be accompanied by other changes, for instance, a re-styling of the journal, a switch to English as the main publication language, re-defining the journal scope, or replacing editorial board members. The current authors plan to conduct such qualitative case studies in a follow-up project.

## Endnotes

1. The Scopus Source List contains information on the OA status at the level of journals, derived from DOAJ. As from 02/01/2018, Scopus contains data on OA status at the level of individual articles, based on an analysis conducted by the Scopus team in collaboration with CrossRef. The Article Level Open Access indicator identifies articles both in Gold OA journals and in hybrid journals. (https://blog.scopus.com/posts/what-s-new-on-scopus-article-level-open-access-indicator-now-at-the-article-level-and-other). ImpactStory's UnPaywall posted on its website https://blog.ourresearch.org/elsevier-data-feed/ already in July 2018 the following message: "We're pleased to announce that Elsevier has become the newest customer of Impactstory's Unpaywall Data Feed, which provides a weekly feed of changes in Unpaywall, our open database of 20 million open access articles. It is uncertain whether UnPayWall-based data were used in Scopus at the date the data for the current study were collected" (End of December 2019). Akbaritabar & Stahlschmidt, (2019) compared Scopus with Unpaywall, and concluded that "Only about 70% of articles and reviews from WOS and Scopus could be matched via a DOI to Unpaywall" (Akbaritabar & Stahlschmidt, 2019, p. 1).